\documentclass[aps,twocolumn,showpacs,superscriptaddress,10pt,longbibliography]{revtex4-1}

\usepackage[colorlinks,bookmarks=false,citecolor=blue,linkcolor=red,urlcolor=blue]{hyperref}

\usepackage{amsmath}
\usepackage{amsfonts}
\usepackage{amssymb}
\usepackage{graphicx}
\usepackage{color}
\usepackage{soul}
\usepackage{xcolor}
\usepackage{footmisc}
\usepackage{braket}
\usepackage{todonotes}

\allowdisplaybreaks

\begin{document}

\title{Stroboscopic aliasing in long-range interacting quantum systems}
%\abstract{sometihng}
\begin{abstract}

We unveil a mechanism for generating oscillations with arbitrary multiplets of the period of a given external drive, in long-range interacting quantum many-particle spin systems. 
These oscillations break discrete time translation symmetry as in time crystals, but they are understood via two intertwined stroboscopic effects similar to the aliasing resulting from video taping a single fast rotating helicopter blade.
 The first effect is similar to a single blade appearing as multiple blades due to a frame rate that is in resonance with the frequency of the helicopter blades' rotation;
 the second is akin to the optical appearance of the helicopter blades moving in reverse direction.
 Analogously to other dynamically stabilized states in interacting quantum many-body systems, this stroboscopic aliasing is robust to detuning and excursions from a chosen set of driving parameters, and it offers a novel route for engineering dynamical $n$-tuplets in long-range quantum simulators, with potential applications to spin squeezing generation and entangled state preparation.
 \end{abstract}

\author{Shane P. Kelly}
\email[Corresponding author:~]{shakelly@uni-mainz.de}
\affiliation{Theoretical Division, Los Alamos National Laboratory, Los Alamos, New Mexico 87545, USA}
\affiliation{Department of Physics and Astronomy, University of California Riverside, Riverside, California 92521, USA}
\affiliation{Institut f\"ur Physik, Johannes Gutenberg Universit\"at Mainz, D-55099 Mainz, Germany}
\author{Eddy Timmermans}
\affiliation{XCP-5, XCP Division, Los Alamos National Laboratory, Los Alamos, New Mexico 87545, USA}
\author{Jamir Marino}
\affiliation{Institut f\"ur Physik, Johannes Gutenberg Universit\"at Mainz, D-55099 Mainz, Germany}
\author{S.-W. Tsai}
\affiliation{Department of Physics and Astronomy, University of California Riverside, Riverside, California 92521, USA}

\maketitle
\emph{Introduction.}~The field of dynamical stabilization has a long tradition tracing back to the Kapitza pendulum in the mid 60s~\cite{kapitza1965dynamical}: a rigid rod can be stabilized in an inverted position by parametrically driving its suspension point with a tuned oscillation amplitude and at high frequency.
The working principle of a dynamically stabilized upside-down pendulum is the building block for realizing periodic motion in atomic physics, plasma physics and in the theory of dynamical control in cybernetical physics. 
Periodic drives are a versatile tool that can be employed to stabilize systems in configurations prohibited at equilibrium. Applications in the quantum domain range from 
 cold atoms to trapped ions~\cite{saito2003dynamically, abdullaev2003controlling,zhang2010localization,hoang4363dynamic,abdullaev2000macroscopic,boukobza2010nonlinear,citro2015dynamical,
lerose2019prethermal}: a drive with large amplitude and fast frequency can stabilize an entire band of excitations, turning the dynamics of a collective mode from a runaway trajectory into a periodic orbit.
In this work, we propose a flexible route to engineer periodic dynamical responses characterized by arbitrary integer fractions of the period of the drive, relevant for a broad class of quantum many-body simulators.

Periodic dynamics in isolated many-particle systems, can be also found in the absence of an external drive. Examples range from quantum 'scars'~\cite{PhysRevLett.122.220603,PhysRevLett.123.147201,turner2018weak,bernien2017probing,PhysRevB.98.235155} to the dynamical confinement of correlations~\cite{lerose2020quasilocalized, robinson2019signatures, PhysRevLett.124.180602, kormos2017real, PhysRevLett.122.150601,PhysRevB.99.180302,tortora2020relaxation} and encompass the role of dynamical symmetries~\cite{PhysRevB.102.041117,PhysRevLett.125.060601,PhysRevB.102.075132,PhysRevB.102.085140,buca2020quantum} in evoking persistent temporal oscillations.
\begin{figure}[h!]
     \includegraphics[width=0.23\textwidth]{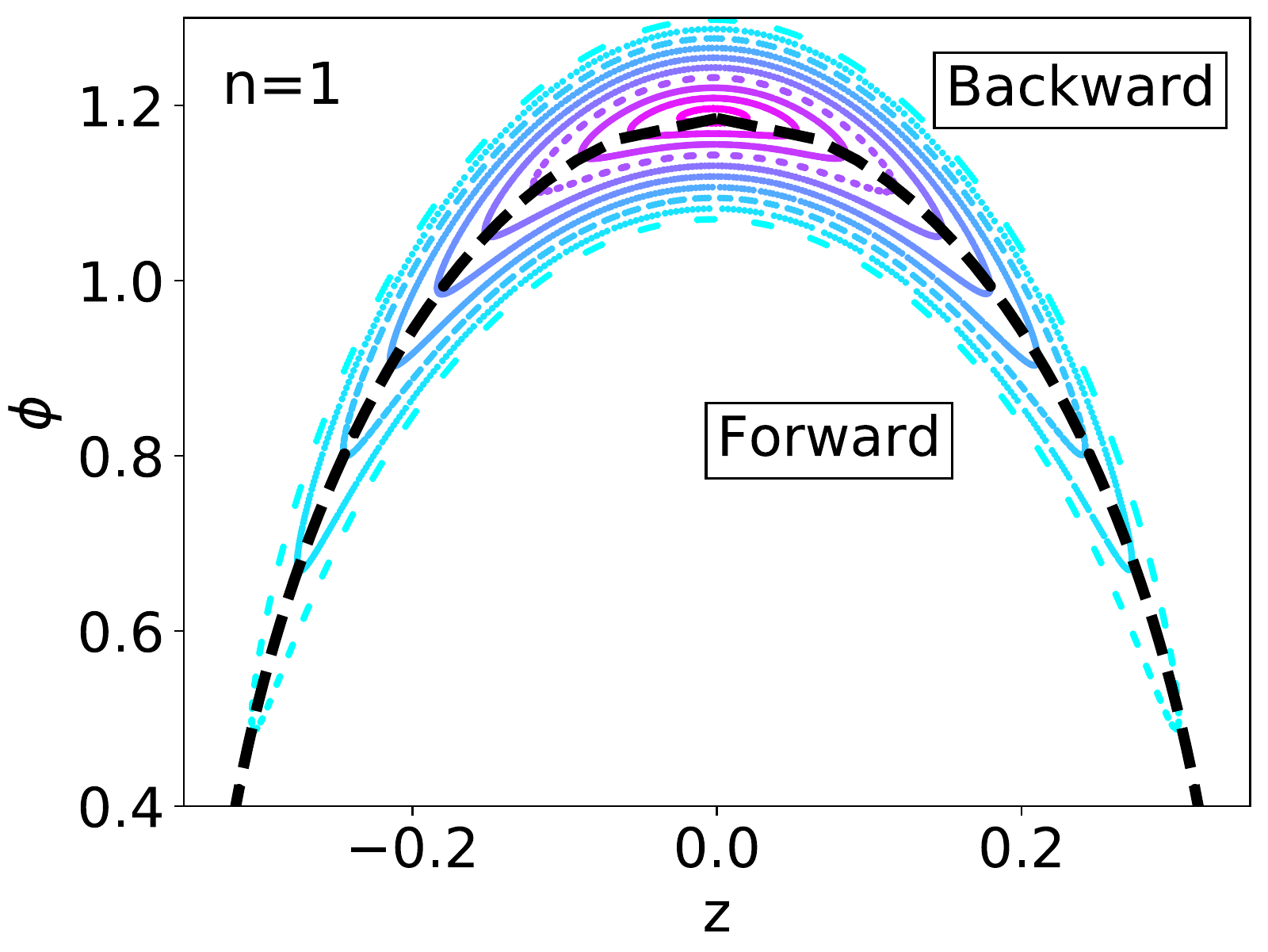}
     \includegraphics[width=0.23\textwidth]{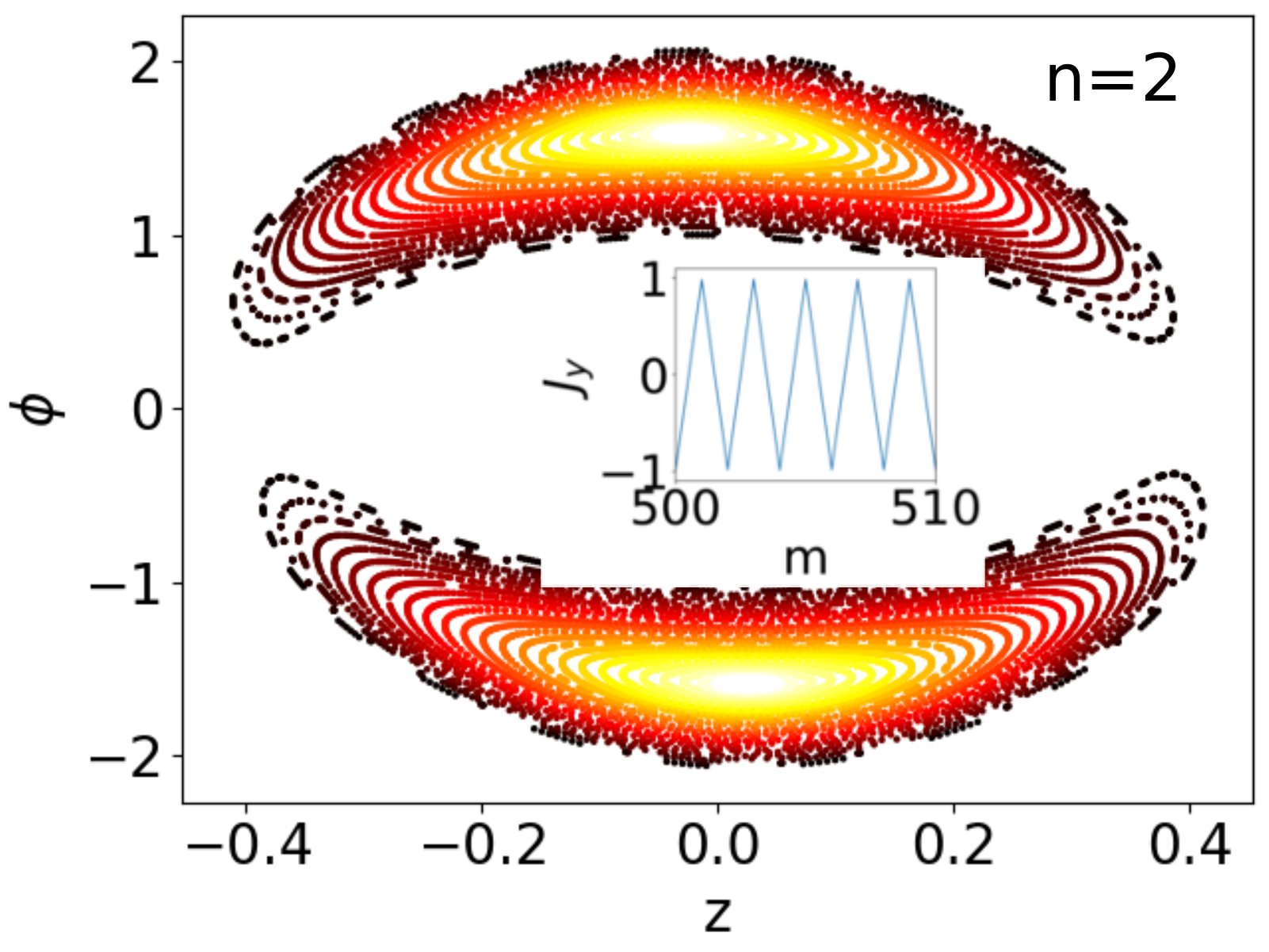}\\
     \includegraphics[width=0.4\textwidth]{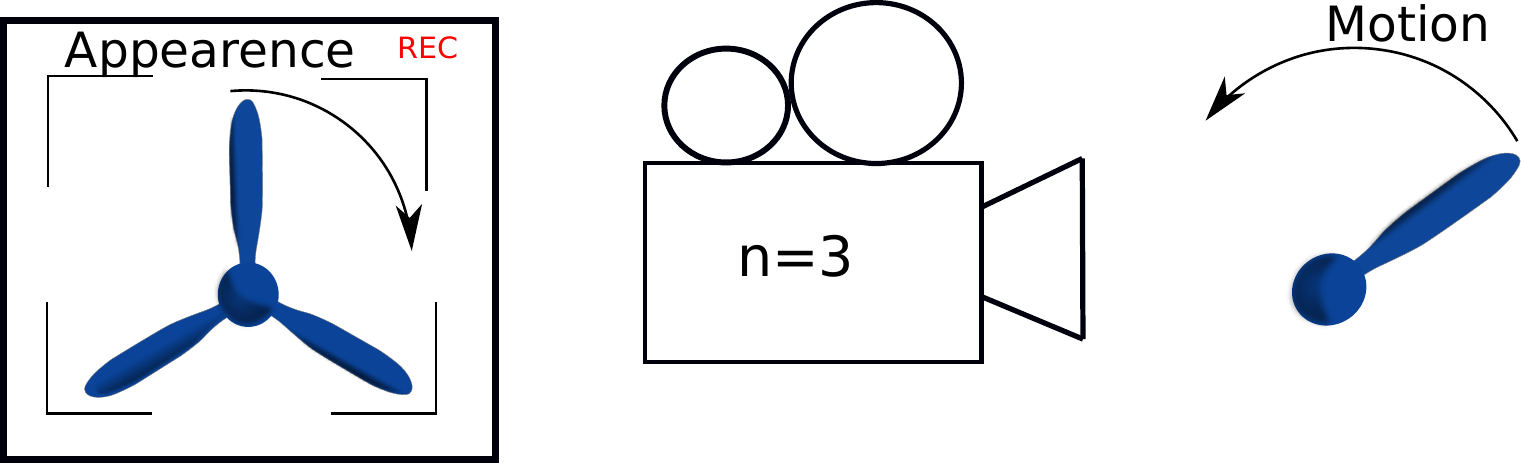}\\
     \caption{[Color Online] The top left figure shows the classical stroboscopic dynamics for an $n=1$ resonance with $(t_1,t_2)=(2.1, 0.005)$. The black line shows the $H_1$ trajectory with period $\tau=nt_1$. In the region labeled ``Forward'' the stroboscopic dynamics appear to move forward along this trajectory (analogous behaviour holds for the region labeled ``Backward''). This apparent reversal of motion is equivalent to the stroboscopic aliasing effect observed when the frame rate of a camera is faster than the rotation rate of a helicopter blade.
The top right figure shows example of an $n=2$ resonance with $(t_1,t_2)=$ $(1.1,0.05)$, and it contains an inset of the exact stroboscopic quantum dynamics that displays the $n=2$ subharmonic response.
The cartoon depicts an example of stroboscopic aliasing effects that occurs when the frame rate of the camera is $n=3-\left|\epsilon\right|$ times the rotation rate of the blade.
 }
     \label{fig:intro0}
 \end{figure}
The quest for time translation breaking in periodically driven quantum systems~\cite{wilczek2012a,watanabe2015} has recently morphed into the search for quantum time crystals~\cite{khemani2019,else2019,sacha2017}. %
A discrete time crystal (DTC) occurs when the discrete time translation symmetry of a periodically driven system is spontaneously broken into a smaller symmetry subgroup. 
One iconic example~\cite{khemani2016b,else2016a} of DTC occurs when the spins of a disordered spin chain are flipped at periodic intervals, and their 
local magnetization oscillates with a period twice the one of the spin flips.
In this model, the stability of the time crystalline behaviour is provided by the extensive set of quasi-local integrals of motion which are characteristic of many-body localized phases occurring at strong disorder~\cite{abanin2015,nandkishore2015a}.
 
Since original experiments in trapped atomic ions and in nitrogen-vacancy centers \cite{zhang2017c,choi2017a}, many other mechanisms for time crystals have been proposed~\cite{li2012space,russomanno2017,gong2018b,iemini2018,barberena2019,zhu2019a,surace2019,pizzi2019a,luitz2020,else2017,abanin2015,tucker2018a} and observed~\cite{pal2018,rovny2018a,rovny2018,rubio-abadal2020,autti2020ac}.
In all of these systems, the periodic dynamics are split into two parts: the natural dynamics of a system that possesses a $Z_n$ symmetry, and a kick process that sequentially switches among the $n$ symmetry sectors.
An $n$-period DTC (or `n-tuplets dynamics') occurs since it takes $n$ of such kick processes to bring the system back to its original configuration~\cite{PhysRevB.96.115127}.

In this work we show how to engineer dynamics with arbitrary $n$-tuplets that are not distinguished by the sectors of a $Z_n$ symmetry. Differently from time crystals, their stability emerges as a cooperative effect between the natural dynamics and the kick process. Subharmonic response with any value of $n$ can be generated provided {that the kick period is in resonance with the $n^{th}$ harmonic of a collective mode, and this collective mode remains stable, though deformed, during the kicked process. This results in stroboscopic dynamics which display $n$ period oscillations between $n$ emergent dynamical fixed points.}

{By considering the kick akin to the sampling performed by a video camera, we identify this subharmonic response as similar to a type of stroboscopic aliasing that occurs when filming a single blade helicopter:
when the helicopter blade is rotating at the $n^{th}$ subharmonic frequency of the camera's frame rate, its video will appear to have $n$ stationary blades. 
Unlike the sampling performed by the camera, the kick acts on the long-range simulator increasing or decreasing the frequency of the system.
This results in another stroboscopic aliasing effect in which the apparent $n$ stationary blade appear to slowly move forward or backwards depending on if the blade frequency was increased or decreased (cf. with Fig.~\ref{fig:intro0}).
We show that for a general class of kicks, both forward and backward aliasing appears and generate a set of $n$ stroboscopic fixed points that stabilize the subharmonic response. 
Stroboscopic aliasing produces also a set of $n$ unstable dynamical fixed points which we argue could be used for generating spin squeezing and entangled states.}

%We study dynamics similar to that done by Russomanno et. al.\cite{russomanno2017}, but we replace the $\pi$ spin rotation around the $x$ axis (spin flip) with an arbitrary rotation around the $x$ axis by a phase $t_2$.

\emph{Stroboscopic Aliasing.}~
We consider a long-range interacting Ising model~\cite{sciolla2013quantum, das2006infinite,dutta2015quantum,tonielli2019orthogonality,lerose2019impact,lerose2018chaotic} in which the interaction strength is periodically kicked $U(m)=(U_1U_2)^m$.
We define $U_a$ as a unitary generated by the following hamiltonian 
\begin{eqnarray}
    H_a=-\sum^N_{k=1} \sigma_{k}^x+\frac{\Lambda_a}{2N^{1-\alpha}}\sum^N_{k,j=1}\frac{\sigma_{k}^{z}\sigma_{j}^{z}}{\left|k-j\right|^{\alpha}},
    \label{eq:ham}
\end{eqnarray}
where $N$ is the number of spin-halfs, $\vec{\sigma}_k$, which live on a one dimensional lattice, and the unitaries are evolved for different times $t_1$ and $t_2$ and for different interaction strengths $\Lambda_1$ and $\Lambda_2$ (i.e. $U_a=e^{it_aH_a}$, with $a=1,2$).
The Kac rescaling factor with $N^{1-\alpha}$ is to ensure the extensivity of the hamiltonian in the thermodynamic limit~\cite{kac1963}.
The subharmonic response emerges when $t_1$ is in resonance with a collective mode of $H_1$ and $t_2\ll t_1$.
Focusing our attention to this limit, we will refer to $U_2$ as the kick.

\begin{figure}[]
     \includegraphics[width=0.22\textwidth]{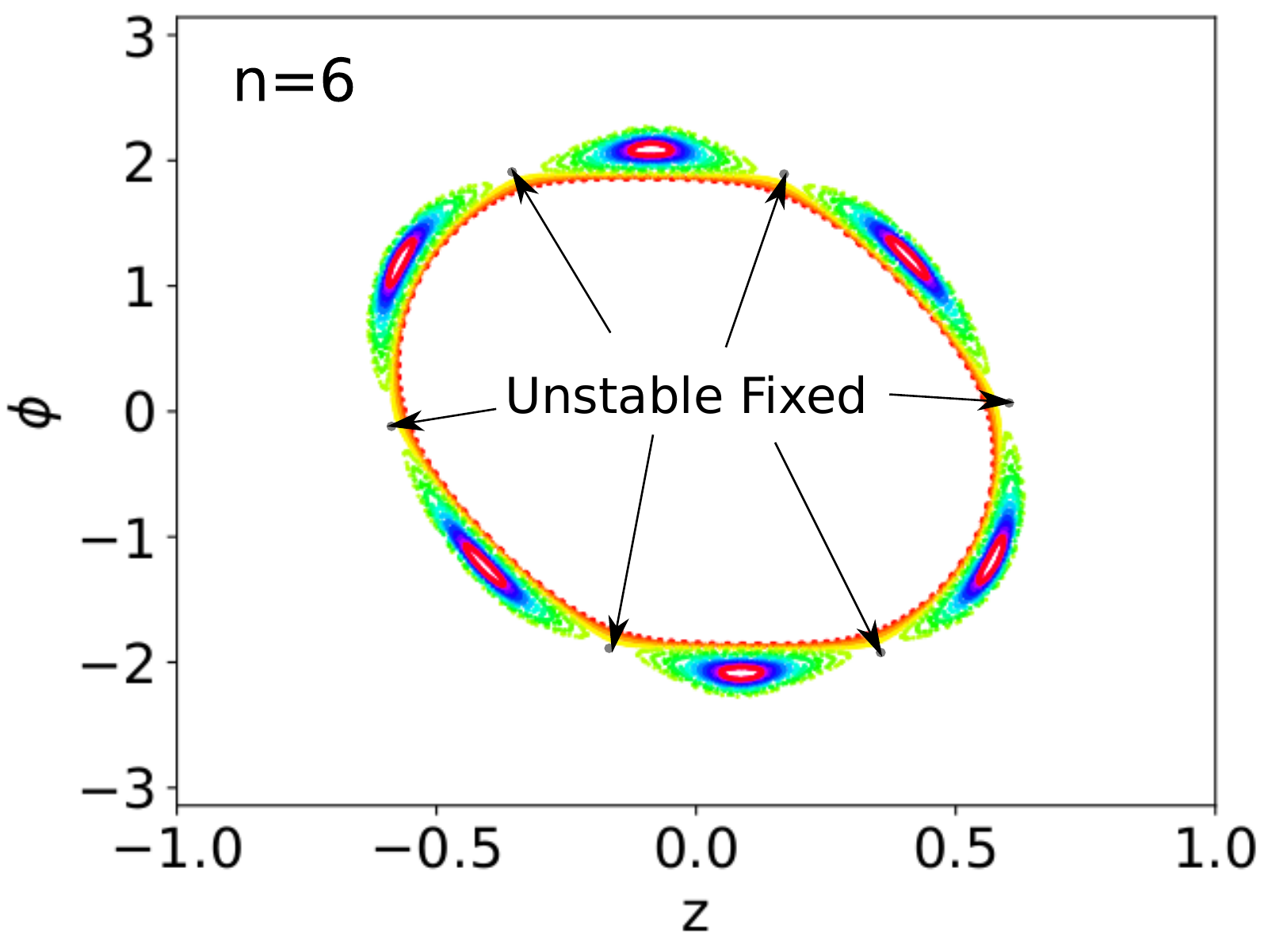}
     \includegraphics[width=0.22\textwidth]{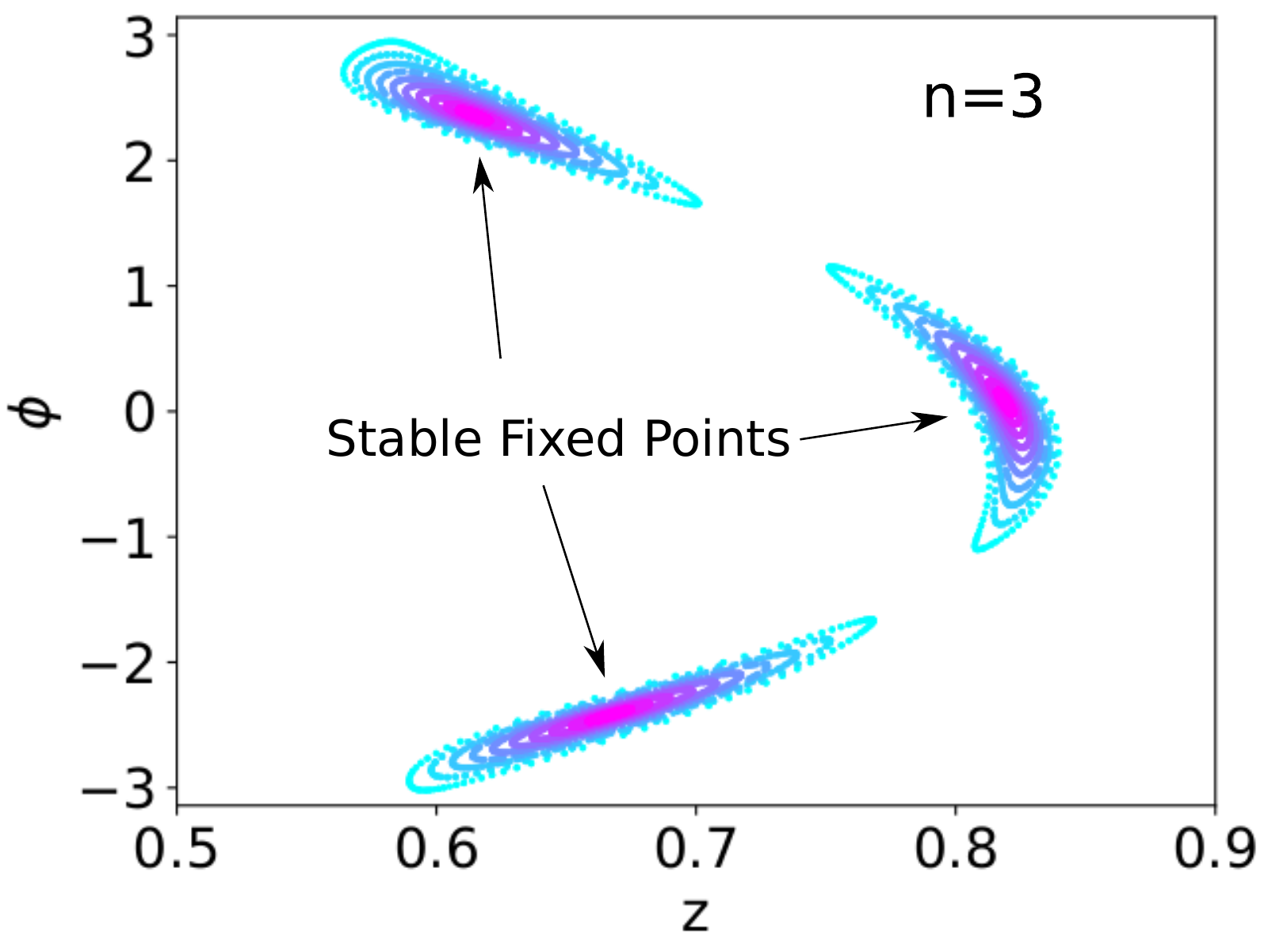}\\
     \includegraphics[width=0.22\textwidth]{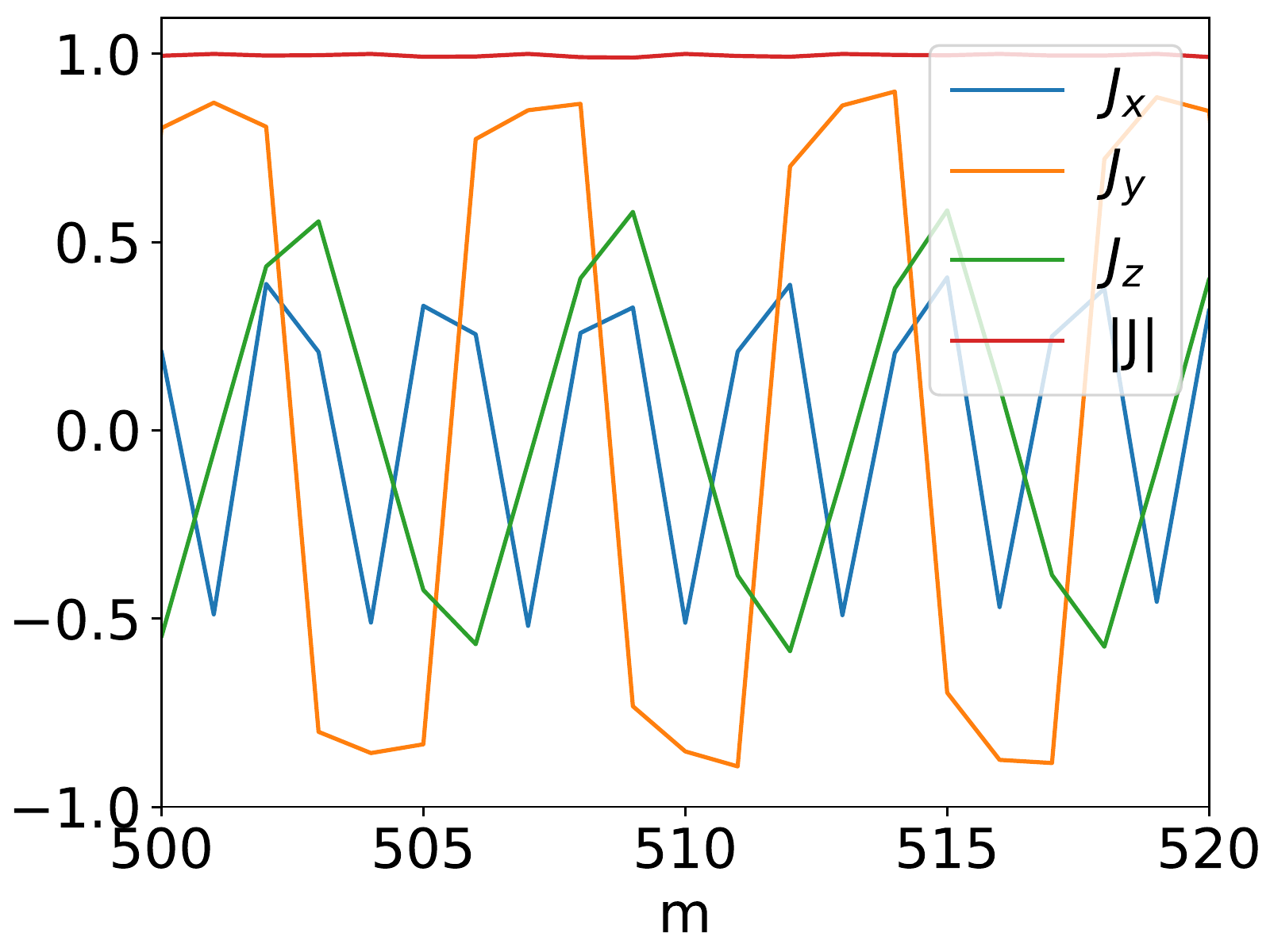}
     \includegraphics[width=0.22\textwidth]{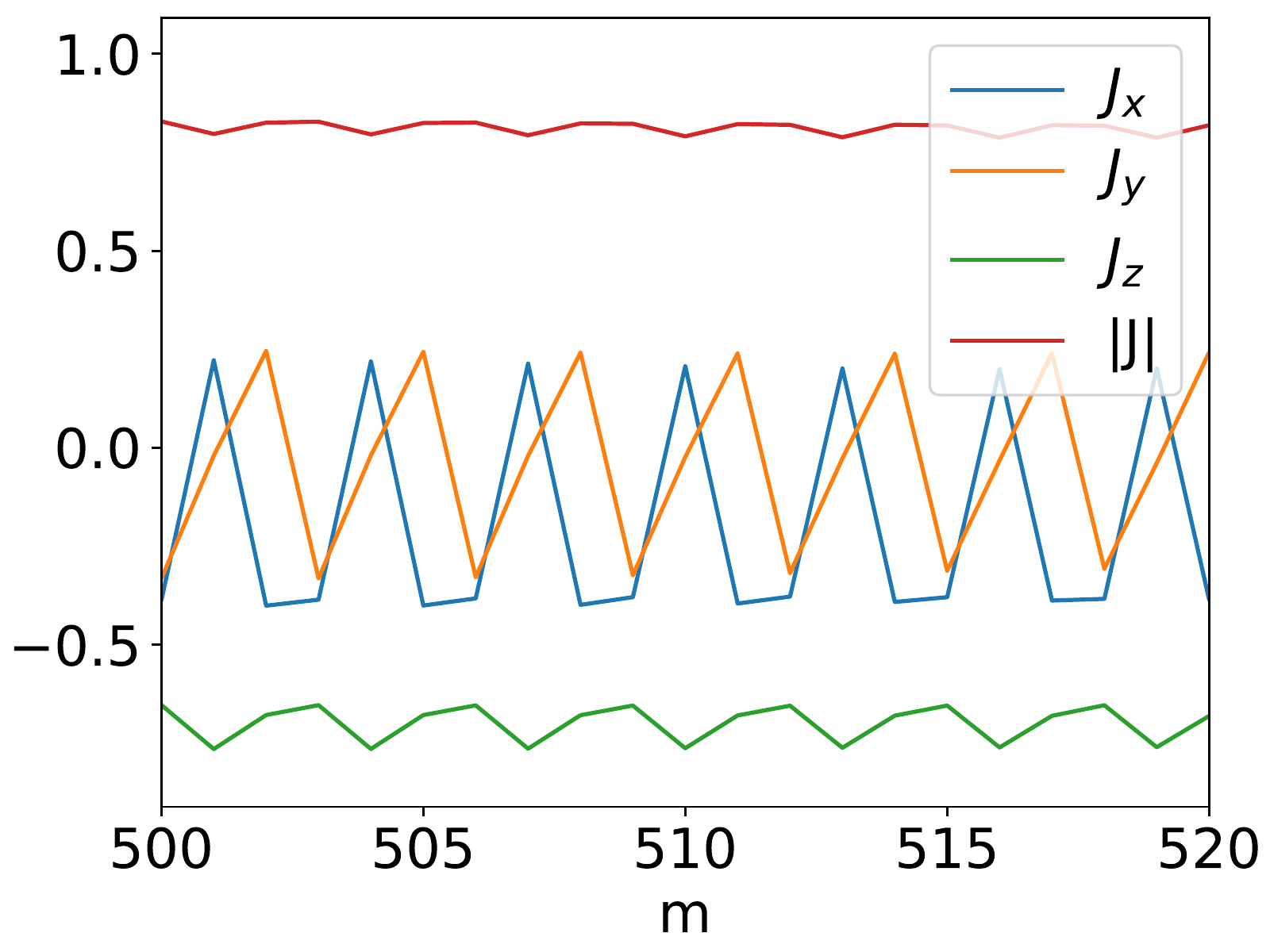}
     \caption{Stroboscopic classical Poincar\'{e} section (top) and exact stroboscopic quantum dynamics (bottom) for $N=500$, $\Lambda_1=10$, $\Lambda_2=0$: with $(t_1,t_2)=$ $(0.35,0.2)$ (left), and $(0.3,0.1)$ (right).
         The color (brightness) in the top plots distinguishes the initial state.
         The top plot depicts the emergent classical fixed points for $n=6$ (left), and $n=3$ (right).
 In the bottom plots, we show the $n=6$ and $n=3$ subharmonic oscillations due to $U_1$ moving between the different emergent fixed points. $\left|J\right|$ is plotted to illustrate that the fixed points stabilize the system against quantum dephasing.}
     \label{fig:intro}
 \end{figure}

The emergent subharmonic response is most clearly explained in the $\alpha=0$ infinite range limit in which the model reduces to the LMG model \cite{glick1965,lipkin1965,meshkov1965}. 
In the large $N$ limit, dynamics reduce to the motion of the collective magnetization $J_\alpha=\frac{1}{N}\sum_i\sigma_{i}^{\alpha}~$\cite{micheli2003}.
The phase space of this collective mode has conjugate variables given by $z$ (the projection of the spin onto the $z$ axis) and by the phase $\phi$ of the spin in the $x$-$y$ plane.
The non-linear classical dynamics of $H_1$ are integrable and can display a separatrix for strong enough $\Lambda_1$.
When $t_1$ and $t_2$ are large, the classical dynamics has a chaotic structure in the same universality class as the standard map \cite{chirikov1979}.
When $t_2$ is small, most of the integrable trajectories of $H_1$ remain unchanged except for when the kick frequency is in resonance with a harmonic of a trajectory of $H_1$; in this case, $t_1\approx \tau/n$, where $\tau$ is the period of a trajectory of $H_1$.

When this condition is met for an integer $n>1$, the dynamics display persistent subharmonic oscillations, and a few instances are shown in Fig.~\ref{fig:intro0} and Fig.~\ref{fig:intro} (with $\Lambda_1=10$ and $\Lambda_2=0$).
To understand why these oscillations occur and to assess their stability, we will first work in the limit $\Lambda_2=0$, and turn our attention to the first plot of Fig.~\ref{fig:intro0} where we have shown a set of $U(m)$ stroboscopic trajectories near an emergent fixed point with a $n=1$ resonance.
There we have also plotted the resonant ($n=1$) trajectory of $H_1$ in black.
Since $t_1=\tau(E)$, $U_1$ completes one period of the trajectory and evolves a spin initialized on this trajectory back to its initial point.
Thus, ignoring for the moment $1/N$ quantum corrections~\cite{raghavan1999}, we can approximate $U_1\approx1$ for initial states on this resonant trajectory.
Similarly, when initial states start on an $H_1$ trajectory with period slightly less than $t_1$, they appear to move slightly forward along the trajectory by a time $t_1-\tau$.
Again, we can approximate $U_1(t_1)\approx U_1(t_1-\tau)$ when $U_1$ acts in this region of phase space.
Similarly when $t_1<\tau$, the state appears to move slightly backwards by a time $\tau-t_1$ and we can approximate $U_1(t_1)\approx U_1^{\dagger}(\tau-t_1)$.
This inspires us to label the trajectories with $\tau<t_1$ as `forward' trajectories and the trajectories with $\tau>t_1$ as `backward' trajectories.
This apparent forward and backward motion is the same stroboscopic aliasing effect that occurs when video taping a helicopter blade with a frame rate similar to the rotation frequency.

 We now consider the action of the $U_2$ kick. For $\Lambda_2=0$, the kick is a $J_x$ rotation, and in the region of phase space shown in the first plot of Fig.~\ref{fig:intro0}, a $J_x$ rotation increases $z$ and keeps $\phi$ approximately constant.
 Therefore, when $z>0$ a spin on a forward trajectory is kicked towards the backward trajectories, while when $z<0$, a spin on a backwards trajectory is kicked towards the forward trajectories.
 Thus, in this region of phase space, the interplay of stroboscopic aliasing and the kick causes the spin to switch back and forth between the forward and backward trajectories and creates a new stroboscopic fixed point.
 For small $t_2$, these non-trivial tori are separated by the perturbed LMG tori by two separatrices that meet at $n$ unstable fixed point (See Fig.~\ref{fig:intro}).

When the resonance condition occurs for $n>1$ a similar description holds up to a few subtleties.
First, $U_1$ only completes a fraction ($1/n$) of a trajectory.
Therefore, we should define the forward and backward trajectories based off the classical trajectories of the unitary, $U'_1=(U_1U_2)^{n-1}U_1$. %\footnote{See SM for a more detailed discussion on this classical approximation}.
In the perturbative limit of small $t_2$, the classical periods and trajectories of $U'_1$ will only be slightly shifted from the LMG trajectories, and we can follow similar arguments as above.
The dynamics defined by $U'(m)=(U'_1U_2)^m$ will then have a similar fixed point structure and trajectories as shown in Fig.~\ref{fig:intro0}, but will only capture the dynamics when looking every $n$ steps of $U$.
Looking at every step, we see that $U$ will shift the fixed point and resonant trajectories of $U'$ to $n$ different $U'$ fixed points in phase space before returning to the original $U'$ fixed point. 
This shows that, at the resonances, there must be $n$ stroboscopic fixed points of the $U'$ dynamics, and this is confirmed in Fig.~\ref{fig:intro}.
Since these are fixed points of the $U'$ dynamics, the $U$ dynamics display a period-$n$ oscillation due to $U$ moving the spin between the $n$ different fixed points of $U'$.
In the analogy to stroboscopic aliasing, this subharmonic response is similar to a filmed single blade helicopter apparently showing multiple $n$ blades when the frame rate $1/t_1$ is $n$ times the frequency of the helicopter $1/\tau$.

\emph{Stability}. 
Unlike the stroboscopic aliasing that occurs while filming helicopters, the stroboscopic aliasing subharmonic response is actively stabilized by the interplay between aliasing and kicking, and it does persist when the drive parameters are slightly detuned. 
First, we discuss the stability of stroboscopic aliasing to the accumulation of quantum fluctuations in the course of long-time dynamics.
In the bare LMG model $H_1$, fluctuations lead to the collapse of periodic oscillations~\cite{PhysRevB.99.045128}, while in the exact\cite{weinberg2017quspin} numerical calculations, we find that such collapse does not occur for the aliasing subharmonic response.
This can be understood in a semiclassical picture where quantum fluctuations are captured by a quantum diffusion process that spreads the wave function along the conservative classical trajectory~\cite{polkovnikov2010phase}.
Collapse of periodic oscillations occurs when the diffusion process reaches a steady state with the wave function completely spread out along the periodic trajectory performed by the classical dynamics.
\begin{figure}[b!]
     \includegraphics[width=0.4\textwidth]{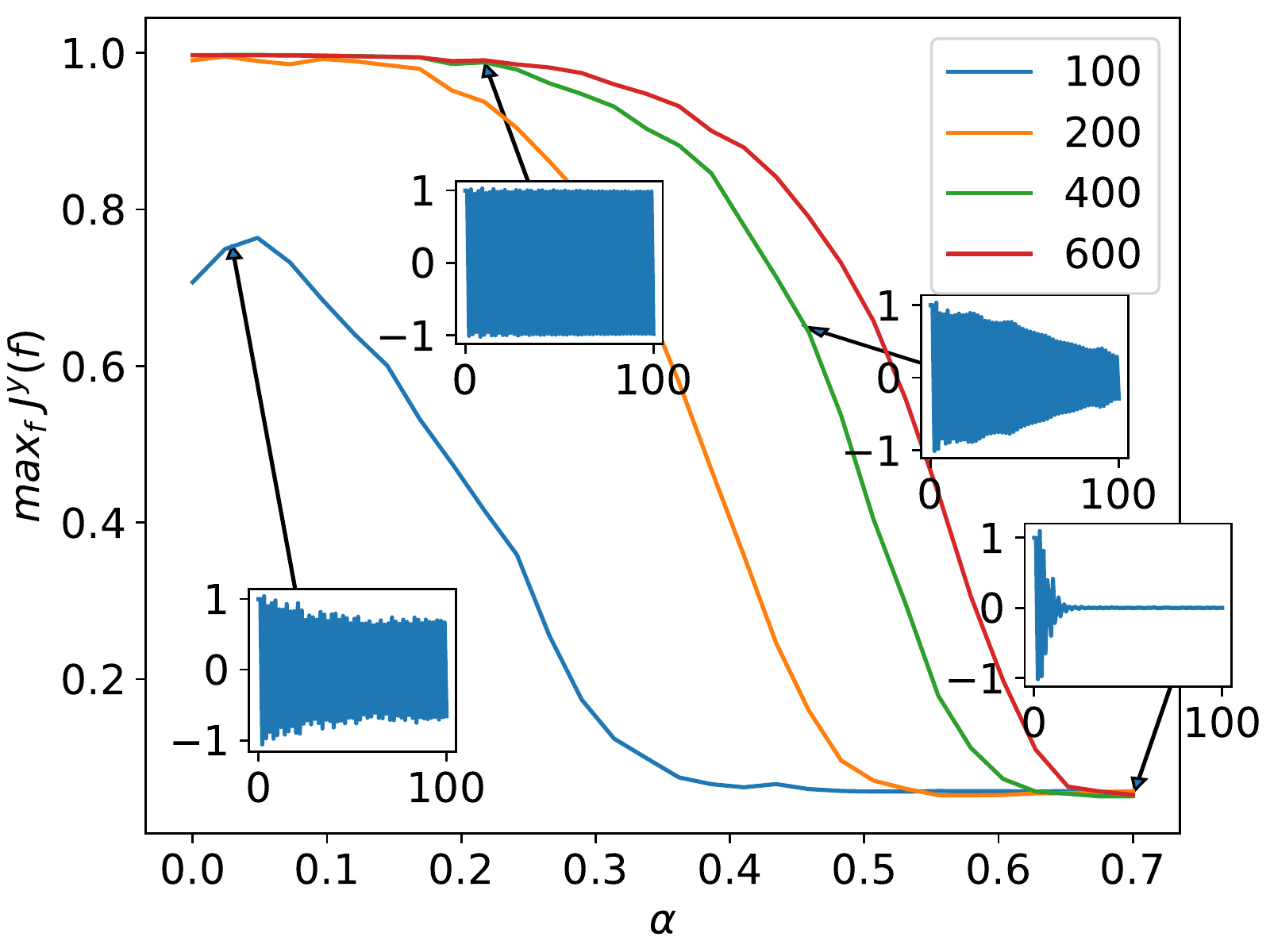}\\
     \includegraphics[width=0.22\textwidth]{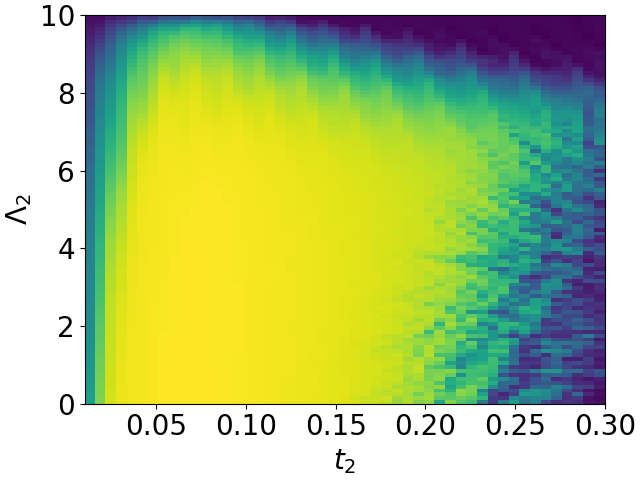}
     \includegraphics[width=0.22\textwidth]{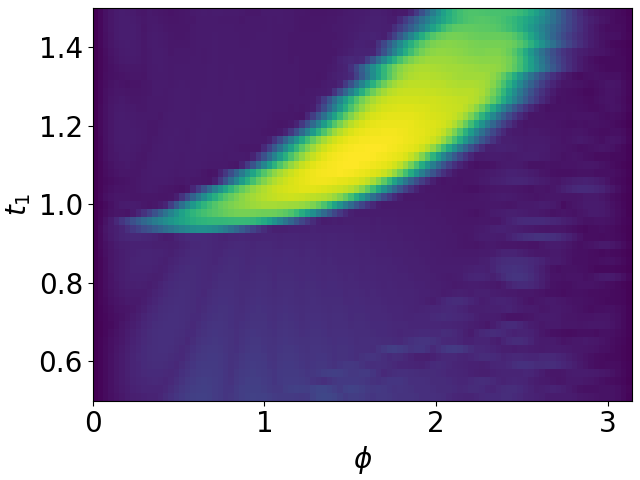}
     \caption{In this plot we demonstrate the stability of the $n=2$ Stroboscopic Aliasing subharmonic oscillations to variation of hamiltonian parameters and many body perturbations. Calculations are done for the hamiltonian~\eqref{eq:ham} in one dimension. The bottom two panels are for $\alpha=0$ and are computed using exact quantum dynamics. They show the order parameter $\max_f J_y(f)$ discussed in the text as a function of $\Lambda_2$, $t_2$(left) and $t_1$ and the initial phase $\phi$ (right). In these plots, the brightest yellow corresponds to $J_y(f)=1$, while the darkest blue to $J_y(f)=0$. The top panel is computed for finite $\alpha$ using DTWA. It shows the same order parameter as a function of $\alpha$, and its insets show $J_y(t)$ at the points indicated by the arrows. 
      }
     \label{fig:zoom_island}
 \end{figure}

For the stroboscopic aliasing subharmonic response, the steady state contains an oscillation that moves the spin between the $n$ dynamical fixed points.
These oscillations remain quantum because the wave function remains localized around these fixed points.
Qualitatively, this is expected by regarding quantum corrections as quantum jumps that move the spin off of its classical trajectory.
In the large $N$ limit, these jumps are exponentially suppressed~\cite{polkovnikov2010phase}, and so they can only move a spin within the well of an emergent fixed point, but not between them.
Thus, we expect that quantum corrections cannot spread the state between the different stable emergent fixed points and that the subharmonic response to be robust to quantum fluctuations.
This is confirmed by the stability of the subharmonic response after $m=500$ oscillations, and the dynamics of $\left|J\right|^2=\sum_\alpha \left<J_\alpha\right>^2$, which shows that spins move along the surface of the Bloch sphere (See Fig.~\ref{fig:intro}).

Therefore, one should expect the stroboscopic aliasing subharmonic response to be stable to variations in $t_a$ and $\Lambda_a$ as long as they only deform the emergent fixed point structure. 
To test the extent of this stability, we focus on the $n=2$ case shown in Fig.~\ref{fig:intro0} and work with an initial state completely polarized along the $J_y$ direction.
As shown in the same figure, the subharmonic response is observed in oscillations of $J_y$ between $1$ and $-1$.
We therefore use the Fourier spectrum, $J_y(f)=\frac{1}{M}\sum_{n=1}^{
}e^{-i f n}J_y(n)$ of the $y$ component of the spin to asses the stability of the stroboscopic aliasing subharmonic response.
When oscillations are stable for long times, the discrete Fourier spectrum, $J_y(f)$ will be singularly peaked around $f=\pi$.
Thus, similar to \cite{pizzi2019a}, we take $\max_f J_y(f)$ as our order parameter for the $n=2$ stroboscopic aliasing oscillation phase.

A phase diagram of this order parameter in the $t_2$ and $\Lambda_2$ parameter space is shown in Fig~\ref{fig:zoom_island}.
The pronounced stability to variation in $\Lambda_2$ reflects the fact that any $U_2$ that connects the forward and backward trajectories in this region of phase space is sufficient to stabilize the fixed point there.
When $t_2$ becomes large, the majority of the resonant trajectories around the fixed points become chaotic and the phase is destroyed.
Fig~\ref{fig:zoom_island} also shows that the phase is stable to variations in $t_1$.
This is because there is a continuum of periods with $\tau=2t_1$ which can be in resonance with $U_1$.

Up to now, we have discussed the limit of $\alpha=0$ in the hamiltonian~\eqref{eq:ham}.
In this case, dynamics are well approximated by the motion of a single large spin, and the evolved states are constrained to a Hilbert space where the spins at different sites are indistinguishable by permutation symmetry.
This Hilbert space has only $N$ states and does not fully reflect the many body nature of a realistic experiment.
Therefore, we study the robustness of the subharmonic response at finite $\alpha$.
We use the Discrete Truncated Wigner Approximation (DTWA) which yields accurate results in long-range interacting models~\cite{schachenmayer2015a,orioli2017,davidson2017semiclassical,acevedo2017exploring,sundar2019analysis,pappalardi2019quantum,khasseh2020discrete}.
DTWA evolves the dynamics according to classical equations of motion, but treats exactly quantum fluctuations in the initial state by sampling over a discrete Wigner distribution~\cite{polkovnikov2010phase}.

We again compute $\max_f J_y(f)$ and the results are shown in Fig.~\ref{fig:zoom_island}.
For $N=100$, quantum diffusion occurs on observable time scales.
As shown in the inset and discussed above for $\alpha=0$, this decreases the amplitude of the subharmonic response but does not result in a complete decay.
For $N=200$, our numerics show that, up to computable time scales, the oscillations are almost perfect up to $\alpha=0.2$ at which the subharmonic response starts to slowly decay.
This indicates that for large values of $\alpha$, many body effects relax the oscillations before quantum diffusion in the collective Hilbert space occurs.
As we increase $N$, this critical $\alpha$ grows to larger values indicating that these many body effects are a finite size effect and are suppressed at large $N$.
While these numerics cannot identify the critical value in the thermodynamic limit, they do show that oscillations are stable for finite $\alpha$, finite $N$ and within observable time scales.
%Therefore, they demonstrate the possibility of experimental observation.

\emph{Generality and Perspectives}. 
 \begin{figure}
     \includegraphics[width=0.4\textwidth]{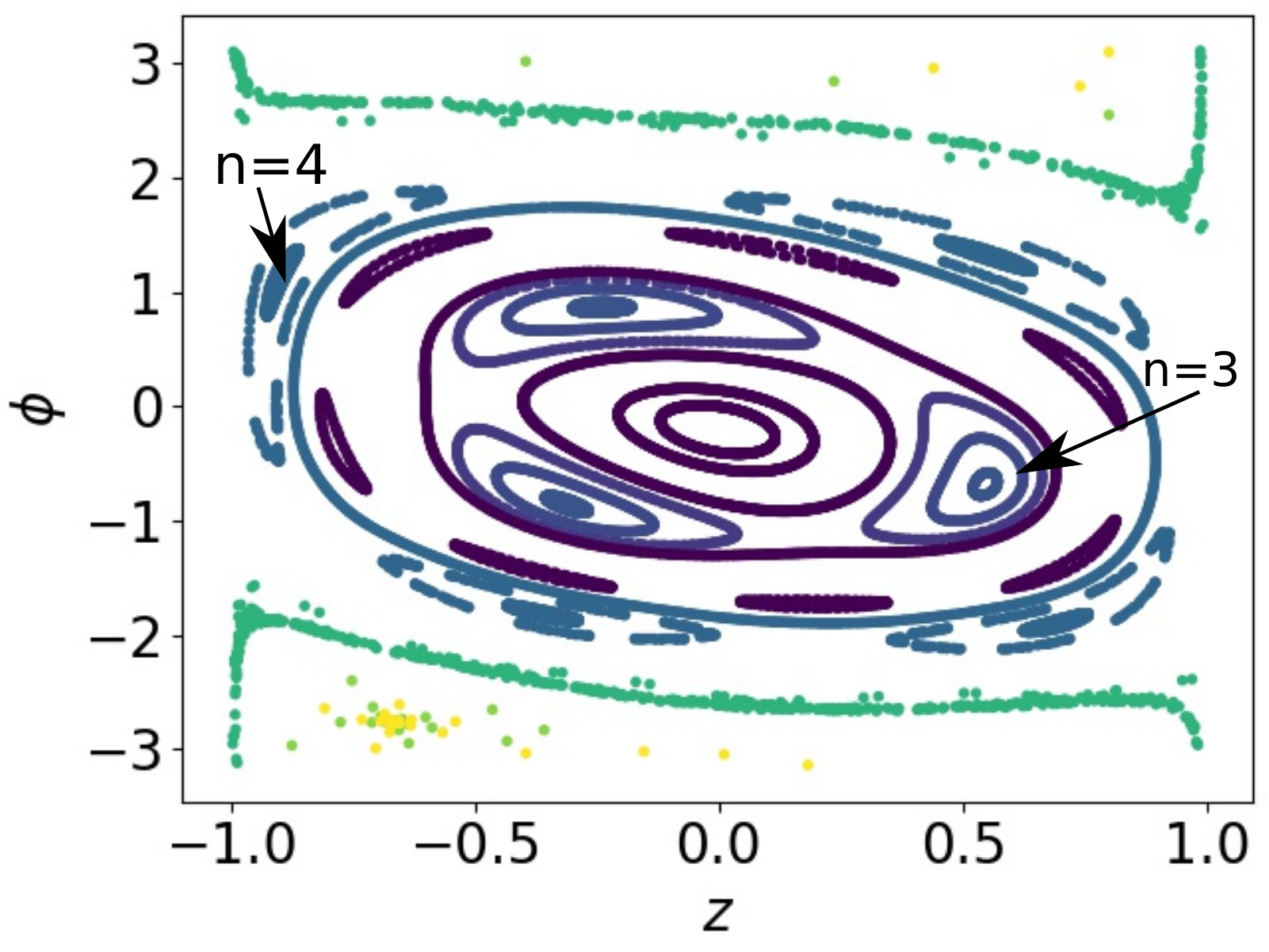}\\
     \caption{Stroboscopic aliasing subharmonic response in the presence of collective spin emission. Depending on initial conditions an $n=4$ or an $n=3$ subharmonic oscillation can occur. Dynamics are computed using the same methods as in~\cite{iemini2018}.}
     \label{fig:disipation}
 \end{figure}
We believe that the stroboscopic aliasing subharmonic response discussed in this work is a general phenomenon provided a few requirements are satisfied.
The collective mode should have only one dominant frequency, otherwise the kick cannot be in resonance with a single period.
Furthermore, the kick must deform the collective mode, although not completely destroy it.
The trajectory of the deformed collective mode should cross the bare trajectory in two points since this will allow for the dynamics of $U'_1=(U_1U_2)^{n-1}U_1$ to cross back and forth across the resonant trajectory.
Notice that these requirements are easily satisfied when the classical phase space of the collective mode is two dimensional because this guarantees regular trajectories with only one frequency.
Despite such required regularity in the collective mode dynamics, integrability is not required as demonstrated by the robustness of the subharmonic response to many body perturbations at finite $\alpha$.
Furthermore, the dynamics of the collective mode is not required to be conservative either.
We demonstrate this aspect by considering the effect of a global spin decay modeled by a Lindblad jump operator proportional to $J^-$, which occurs naturally in cavity QED experiments~\cite{PhysRevLett.122.010405, muniz2020exploring, PhysRevA.99.051803, PhysRevX.9.041011}.
For $\kappa=0.5$ the model has a limit cycle for initial states polarized close to $J_x=-1$~\cite{iemini2018} during its natural evolution, $t_1$.
Choosing $t_1$ to be in resonance with the period of these collective modes, we are able to find a subharmonic response and have plotted examples for $n=4$ and $n=3$ in Fig.~\ref{fig:disipation}.

To conclude, we remark that the stroboscopic aliasing effects discussed so far should be observable in experiments.
The hamiltonian~\eqref{eq:ham} is used to describe trapped ion experiments~\cite{britton2012,zhang2017observation} in which the transverse field is easily controlled and can be employed to implement the kicks of $\Lambda_i$.
Furthermore, the emergent unstable fixed points could also be used to create squeezing or more general entangled states in a way similar to the bare unstable fixed points of $H_1$.
Similar to Refs.~\cite{mahmud2005,micheli2003,strobel2014,kelly2019} such fixed points have two stable directions and two unstable directions.
A quantum state initialized on the unstable fixed point, compresses in the two stable directions and expands in the two unstable direction creating, on short times, a squeezed state.
At longer times, the state is stretched further apart and no longer resembles a squeezed state, yet it might show non-gaussian entanglement with properties controlled by the shape of the separatrix~\cite{strobel2014}.
Since separatrices in the stroboscopic aliasing discussed here, have different topologies, they can open opportunities to generate new classes of entangled states in trapped ions simulators or in ultracold atoms experiments~\cite{micheli2003, kelly2019}, potentially with novel metrological uses.
Finally, studying the critical properties of the transition away from the stroboscopic aliasing response, and analyzing its interplay with quantum fluctuations~\cite{PhysRevLett.120.130603,sartori2015spin} remains an interesting future direction of research.

\begin{acknowledgments}
    \textbf{Acknowledgments:} S. P. K. would like to acknowledge stimulating discussions with Levent Subasi and David Campbell. S. P. K. acknowledges financial support from the UC Office of the President through the UC Laboratory Fees Research Program, Award Number LGF-17- 476883.
    S. P. K. and J. M. acknowledge support by the Dynamics and Topology Centre funded by the State of Rhineland Palatinate.
    S. W. T. acknowledge support by National Science Foundation (NSF) RAISE TAQS (award no. 1839153).
The research of E. T. in the work presented in this manuscript was supported by the Laboratory Directed Research and Development program of Los Alamos National Laboratory under project number 20180045DR.
\end{acknowledgments}

 \bibliography{StrobasticAliasing}

 \end{document}